\documentclass[twocolumn,showpacs,preprintnumbers,amsmath,amssymb,prb]{revtex4}
\usepackage{graphicx}
\usepackage[colorlinks=true,linkcolor=blue,citecolor=blue]{hyperref}
\usepackage{color}
\usepackage{bbm}

\begin{document}

\title{Kosterlitz-Thouless phase transition and reentrance in an anisotropic 3-state Potts model on the generalized Kagom\'e lattice }
\author{Yang Zhao, Wei Li, Bin Xi, Zhe Zhang, Xin Yan, Shi-Ju Ran, Tao Liu, Gang Su}
\email[Author to whom correspondence should be addressed. ]{Email:
gsu@ucas.ac.cn}
 \affiliation{Theoretical Condensed Matter Physics and Computational
Materials Physics Laboratory, School of Physics, University of Chinese Academy
of Sciences, P. O. Box 4588, Beijing 100049, China}

\begin{abstract}
The unusual reentrant phenomenon is observed in the anisotropic 3-state Potts model on a generalized Kagom\'e lattice. By employing the linearized tensor renormalization group method, we find that the reentrance can appear in the region not only under a partial ordered phase as commonly known but also a phase without a local order parameter, which is uncovered to fall into the universality of the Kosterlitz-Thouless (KT) type. The region of the reentrance depends strongly on the ratios of the next nearest couplings $\alpha=J_{2}/|J_{1}|$ and $\beta=J_{3}/|J_{1}|$. The phase diagrams in the plane of temperature versus $\beta$ for different $\alpha$ are obtained. Through massive calculations, it is also revealed that the quasi-entanglement entropy can be used to accurately detect the KT transition temperature.
\end{abstract}

\maketitle

\section{Introduction}

Two-dimensional (2D) frustrated classical lattices possess many interesting physical properties (see, e.g. Refs. [\onlinecite{cent squre, honey,reent,weili,book,xy,classical con,Chenqn}]. One of them is the so-called reentrant phenomenon \cite{cent squre,reent,xy,Potts}, which is defined as the occurrence of a disordered phase (usually a paramagnetic phase) in the region under an ordered or a partial ordered phase on the temperature scale. This disordered phase is called the reentrant phase that has been detected experimentally in spin glasses \cite{expe} and studied theoretically in several exactly solvable 2D frustrated Ising systems on lattices such as the centered square lattice \cite{cent squre}, generalized Kagom\'e lattice \cite{reent}, centered honeycomb lattice \cite{honey} and other three-dimensional lattices \cite{book}. Up to now, the reason for the occurrence of the reentrant phase is still under debated.

Some conjectures \cite{cent squre,reent,book} have been proposed to understand the reentrant phenomenon, in which the essential is that the reentrant phase is probably caused by frustrations and in the ground state, there should be at least one partial order phase adjacent to an ordered phase or another partial ordered phase between which the reentrant phase can appear at finite temperature. Above the reentrant phase, there is usually a partial ordered phase whose disordered sublattices supplement the entropy which is lost due to the formation of the ordered sublattice. In addition, other factors such as the coordinate numbers \cite{book} of the sites on the disordered sublattice and the freedom of the on-site spin \cite{Potts} (e.g. the value of q in the Potts model) may also have effects on the reentrant phenomenon.

Alternatively, to explore the incentives of the reentrant phenomenon, the q-state Potts model \cite{WFY} can offer some clues. Potts model is the generalization of Ising model by extending the on-site freedom from $q$ = 2 to $q > 2$. The q-state Potts model does not have exact solutions. For some lattices such as the piled-up domino model (frustrated Villain lattice) \cite{Potts}, the Potts model possesses a reentrant phase that could maintain in the region under one partial ordered phase when 1 $\leq q \leq$ 4 (except for $q$ = 2), which indicates that the occurrence of reentrant phenomenon has close connection to the value of $q$.

\begin{figure}[tbp]
\includegraphics[width = 0.7\linewidth,clip]{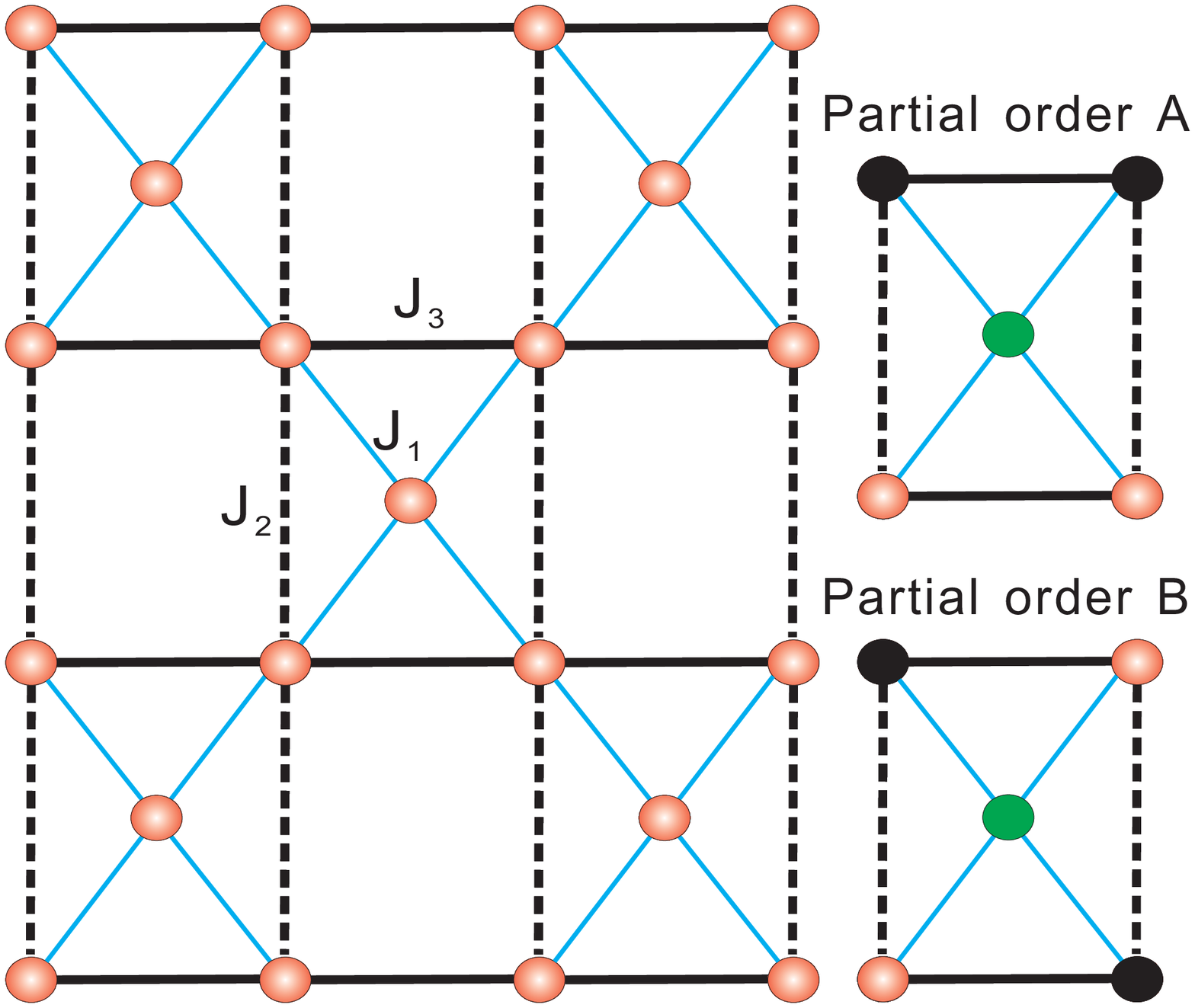}
\caption{(Color online)  The geometric structure (left) and partial order configurations (right) of the generalized Kagom\'e lattice. The thin blue line represents ferromagnetic coupling $J_{1}$, the thick solid line ($J_{3}$) and slash black line ($J_{2}$) could be either ferromagnetic or antiferromagnetic coupling. The black and red dots in partial order A and B configurations indicate different directions of the on-site spin and the centered green site is a free spin. }
\label{Structure}
\end{figure}

In this work, we shall focus on the anisotropic 3-state Potts model on a generalized Kagom\'e lattice, as shown in Fig. \ref{Structure},  which contains three different couplings: $J_{1}$ is the diagonal coupling and is presumed to be ferromagnetic (F), $J_{2}$ and $J_{3}$ are couplings along the vertical and horizontal directions, respectively, both of which can be either antiferromagnetic (AF) or F.  For the Ising case ($q=2$), in the ground states there exist two partial disordered configurations A and B, as shown in Fig. \ref{Structure}. Configuration A  corresponds to the case of $J_{3}<0$ (F) and $J_{2}>0$ (AF), and B is the case of $J_{2}$, $J_{3}$$>$0. The spins on the central site in both cases are in the free states, therefore the order is defined as partial order or partial disorder. By the exact solution the reentrant phase is found between the F phase and the partial ordered phase with configuration A \cite{reent}, which is induced by frustrations.
It is interesting to ask whether there is any reentrant phenomenon in the present anisotropic 3-state Potts model on the generalized Kagom\'e lattice, in which $J_{2}$ is set to be negative and $J_{3}$ positive with $J_{1}$ = -1 (F). Such a mixed (with AF and F interactions) Potts model cannot be solved exactly, and we perform numerical simulations to obtain the specific heat, susceptibility, correlation length, and entanglement entropy of the system to determine the  phase diagram precisely. It is discovered that although the thermal fluctuation depresses any partial order with configuration A and B in the ground state, our numerical results strongly support that there still exists a reentrance but with the Kosterlitz-Thouless (KT) type phase transition at certain values of $J_{2}/|J_{1}|$ and $J_{3}/|J_{1}|$. In addition, it is found that the reentrance disappears when q $>$ 3.

This paper is organized as follows. In Sec. II, the model Hamiltonian and the partition function are defined, and the tensor network representation of this model is given. In Sec. III, the specific heat, susceptibility and the entanglement entropy are calculated, and the KT type phase transition is discussed. In Sec. IV, by examining the singularities of the thermodynamic quantities, the phase diagrams in $\beta-T$ plane are presented for $\alpha=-1$, $0<\alpha<1$ and $\alpha>1$, respectively, and the reentrant phenomena are observed. Finally, a summary is given.

\section{Model and Partition Function }

The Hamiltonian of the anisotropic q-state Potts model on the generalized Kagom\'e lattice reads
\begin{equation}\label{ham}
  H = J_1\sum_{\langle i,j\rangle}\delta_{\sigma_{i}\sigma_{j}}+J_2\sum_{\langle i,j\rangle}\delta_{\sigma_{i}\sigma_{j}}+J_3\sum_{\langle i,j\rangle}\delta_{\sigma_{i}\sigma_{j}},
\end{equation}
where $J_i$ (i = 1, 2, 3) are coupling constants, $\delta_{\sigma_i\sigma_j}$ is the Kronecker symbol with $\sigma_i=1, 2, \cdots q$. In the following we shall focus on q=3. For clarity, we define $\alpha=J_{2}/|J_{1}|$, and $\beta=J_{3}/|J_{1}|$.

By means of the Trotter-Suzuki decomposition the partition function of the system can be written as
\begin{equation}\label{partion}
  Z = Tr e^{-H/T}= Tr{\prod e^{-\epsilon h}},
\end{equation}
\begin{eqnarray}
 h = J_1(\delta_{\sigma\sigma_{1}}+\delta_{\sigma\sigma_{2}}+\delta_{\sigma\sigma_{3}}+\delta_{\sigma\sigma_{4}})\\ \nonumber
 +J_2(\delta_{\sigma_{1}\sigma_{3}}+\delta_{\sigma_{2}\sigma_{4}})
 +J_3(\delta_{\sigma_{1}\sigma_{2}}+\delta_{\sigma_{3}\sigma_{4}}),
 \end{eqnarray}
where $T$ is temperature, the Boltzmann's constant $k_B$ is taken as unity,  and $\epsilon$ is the Trotter slice.

To study this present anisotropic 3-state Potts model, we employ the linearized tensor renormalization group (LTRG) method \cite{Ltrg}, a recently developed numerical algorithm for calculating the thermodynamic properties of the low-dimensional quantum lattice systems with highly accuracy and efficiency, which has been successfully applied to a few one-dimensional quantum spin lattice models \cite{yan,liu}. In the tensor network representation of the partition function (Fig. \ref{Tensor}),
$e^{-\epsilon h}$ can be seen as a fourth-order tensor $\mathbf{T}$ as shown in Fig. \ref{Tensor}(b), and the partition function $Z$ can be represented as a tensor network in Fig. \ref{Tensor}(c).
During the LTRG calculations, we keep the bond dimension cut-off $D_{c}$ of the tensor network at least 60 and the truncation error is less than $10^{-7}$.

\section{Thermodynamic quantities and Kosterlitz-Thouless phase transition}

Kosterlitz-Thouless phase transition in the mixed Potts model \cite{KT, fafpotts} is characterized by the divergence of the correlation length $\xi$ near the critical temperature $T_{c,KT}$ of the form
\begin{equation}\label{corrl}
  \xi \sim \textrm{e}^{\frac{const.}{\sqrt{T-T_{c,KT}}}},
\end{equation}
 while the specific heat shows no divergence and has a broad peak near $T_{c,KT}$. Below $T_{c,KT}$, there is a phase (coined as floating phase) with an algebraically decaying correlation function. Such kind of phase was explained as the resemble of vortices that are closely bound in pairs in 2D XY models \cite{KT}. In the present 3-state Potts model, we also found such a phase and in the following we will follow Ref. [\onlinecite{fafpotts}] to call it the floating phase.

\begin{figure}[tbp]
\includegraphics[width = 0.8\linewidth]{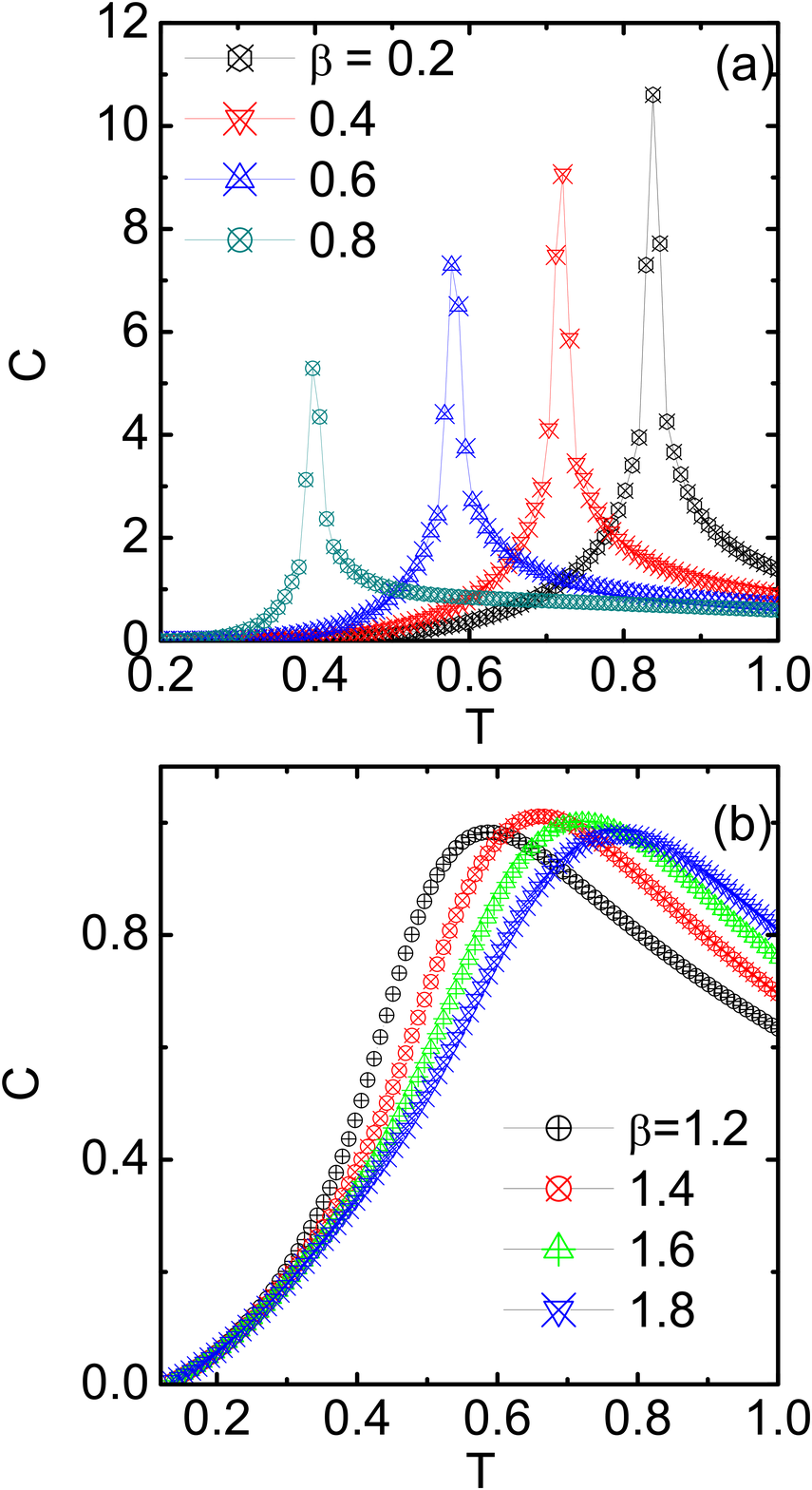}
\caption{(Color online) Temperature dependence of the specific heat of the anisotropic 3-state Potts model on the generalized Kagom\'e lattice with $\alpha$ = -1. (a) $\beta$ $<$ 1, and (b) $\beta$ $>$ 1. }
\label{Capacity}
\end{figure}

\begin{figure}[tbp]
\includegraphics[width = 0.8\linewidth,clip]{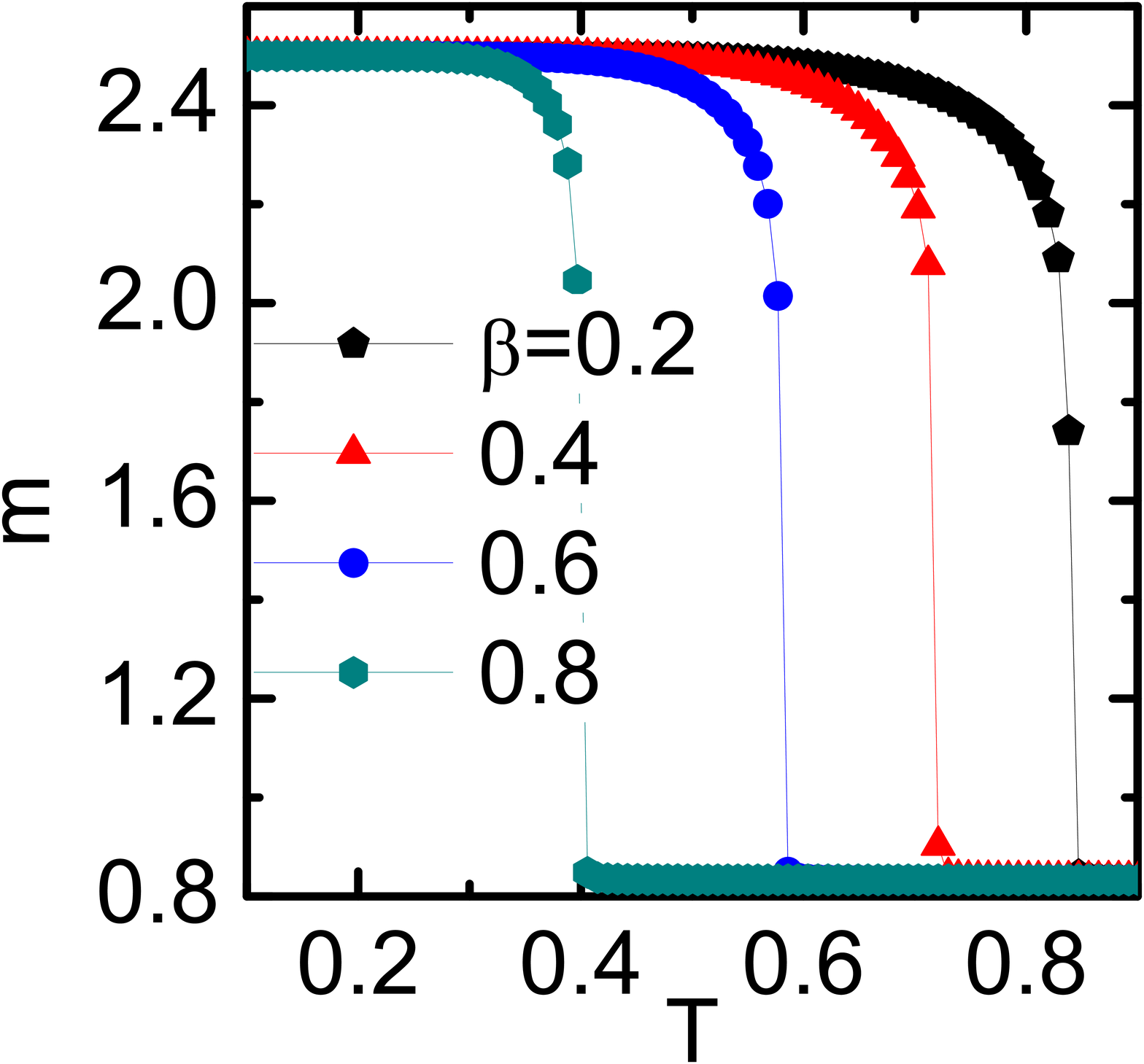}
\caption{(Color online)  Temperature dependence of the magnetization per site of the anisotropic 3-state Potts model on the generalized Kagom\'e lattice with $\alpha$ = -1 and $\beta < 1$.}
\label{M_T}
\end{figure}

Now let us first explore the thermodynamic properties of the generalized Kagom\'e Potts model with parameters $\alpha =J_{2}/|J_{1}|$ = -1 and $\beta= J_{3}/|J_{1}|>0$. The temperature dependence of the specific heat $C$ for different $\beta$ is obtained, as shown in Fig. \ref{Capacity}. It is seen that, for $\beta$ $<$ 1, the specific heat has sharp peaks at critical temperatures [Fig. \ref{Capacity}(a)], which indicate that the second-order phase transitions may take place in this case. This could be easily understood that when $\beta < 1$ the ferromagnetic couplings $J_{2}$ and $J_{1}$ are dominant, and for a given $\beta<1$ the system may undergo a phase transition from ferromagnetic phase to paramagnetic phase, which will also be confirmed in calculations of magnetization (see Fig. \ref{M_T}).
It is observed that the critical temperature $T_{c}$ decreases with increasing $\beta$ when $\beta$ $<$ 1. For $\beta>1$, the specific heat shows broad peaks at low temperatures, and no divergence is found, displaying that no thermodynamic phase transitions can happen in this case. However, this does not imply that the topological phase transition like KT phase transition is unlikely. In fact, it is the case, as seen below.

\begin{figure}[tbp]
\includegraphics[width = 0.8\linewidth,clip]{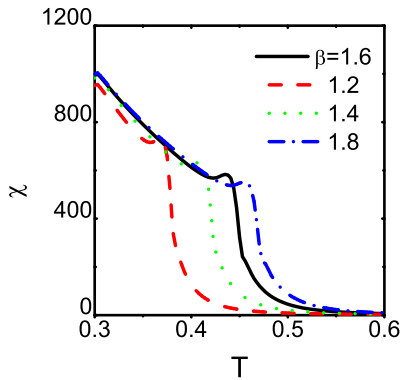}
\caption{(Color online) Temperature dependence of the susceptibility of the anisotropic 3-state Potts model on the generalized Kagom\'e lattice with $\alpha$ = -1 and $\beta > 1$.}
\label{sus}
\end{figure}

To understand further the thermodynamic behavior of this model, we come to look at the magnetization per site defined as
 \begin{equation}\label{m uni}
    m = \frac{1}{N}\sum_{i}\langle \sigma_i \rangle,
\end{equation}
 where N is the number of lattice sites, and $\langle...\rangle$ denotes the thermal average. Note that in this definition $m$ describes the mean value of magnetization on each site whose value equals $\frac{5}{6}$ when the lattice is in disordered state. Figure \ref{M_T} presents the temperature dependence of magnetization per site at $\alpha=-1$ for different $\beta<1$. The sharp changes of $m$ at critical temperatures can be seen, showing that for a given $\beta<1$ the system indeed has a thermodynamic order-disorder phase transition with increasing temperature, being consistent with the results of the specific heat [Fig. \ref{Capacity}(a)]. For $\beta>1$, $m=0$ at finite temperature, showing that in this case no long-range order appears. To examine if there is a topological phase transition in the case of $\beta>1$, we calculated the susceptibility $\chi$ of the system defined by
\begin{equation}\label{m sus}
    \chi = \frac{\partial m}{\partial h},
\end{equation}
where $h$ is the uniform magnetic field. The results are given in Fig. \ref{sus}, where, for
 different values of $\beta>1$, there are singularities (kinks) at certain temperatures, implying that there might be a kind of non-thermodynamic phase transition occurs.

\begin{figure}[tbp]
\includegraphics[width = 0.8\linewidth,clip]{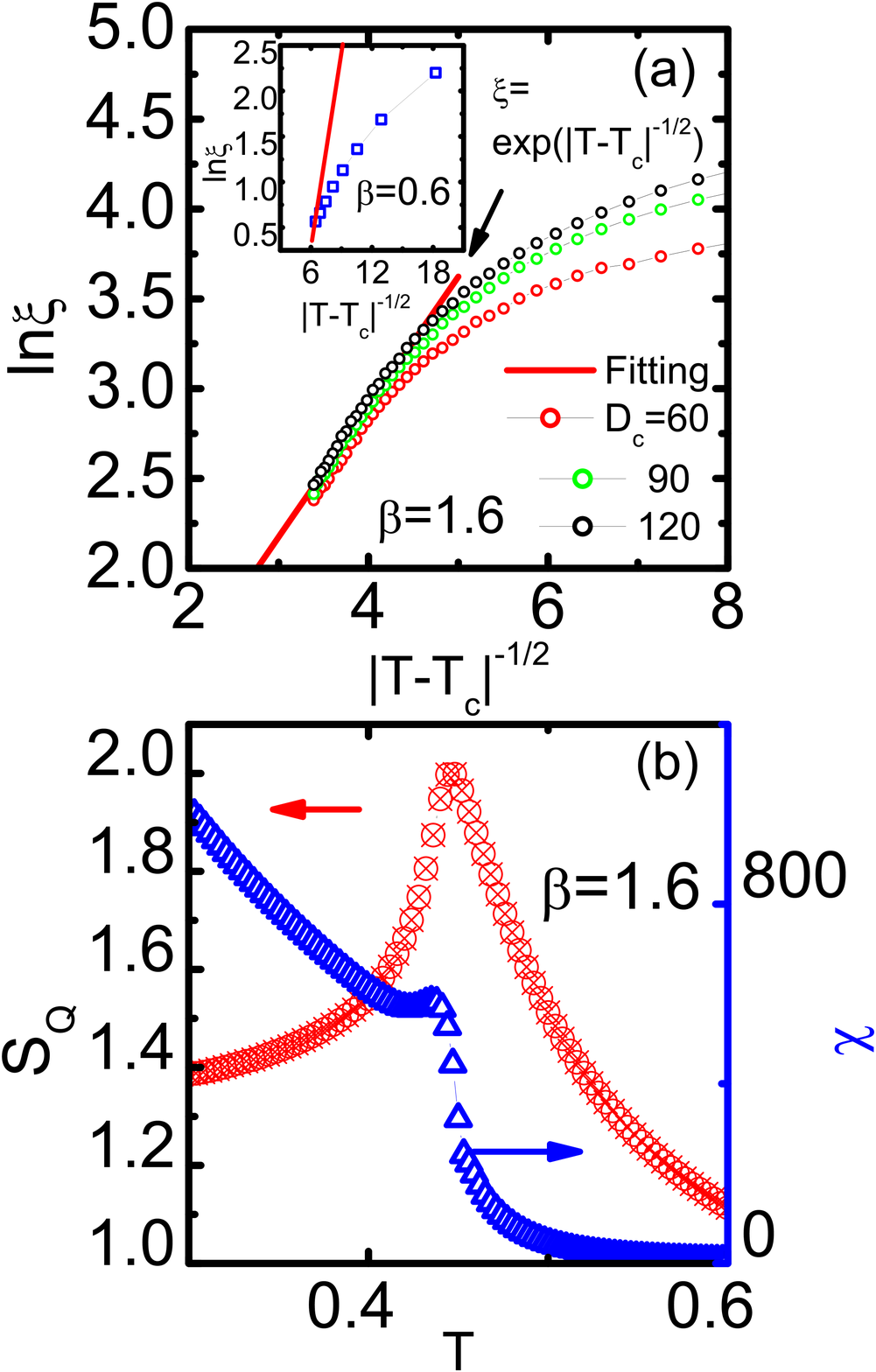}
\caption{(Color online) (a) Correlation length of the anisotropic 3-state Potts model on the generalized Kagom\'e lattice with $\alpha$ = -1 and $\beta > 1$, and the inset is for $\beta < 1$. (b) Quasi entanglement entropy and susceptibility of this model with $\beta=1.6$.}
\label{c and e}
\end{figure}

 To confirm such topological phase transition is of the KT type, we investigate whether the correlation length $\xi$ has the form of Eq. (4). Following the lines introduced in Ref. \onlinecite{corr}, we obtained the results of correlation length versus temperature, as shown in Fig. \ref{c and e} (a) . Apparently, when $\beta>1$ near the transition temperature $T_{c,KT}$, the correlation length $\xi$ shows the same behavior as that given in Eq. (\ref{corrl}), where a slight deviation could be modified by increasing the bond dimension $D_c$. This result supports that the phase transition for $\beta>1$ is indeed of the KT type. For $\beta<1$, the correlation length $\xi$ reveals a distinct behavior from the case of $\beta>1$, as shown in the inset of Fig. \ref{c and e} (a), consistent with the previous observation that the phase transition occurring in the case of $\beta<1$ is thermodynamic. Besides, by fitting the curve, we obtain the critical exponent $\nu = 0.820$ for $\beta<1$ in this model, which is close to the conjectured value $\frac{5}{6}$ (Ref. [\onlinecite{WFY}]).

\begin{figure}[tbp]
\includegraphics[width = 0.8\linewidth]{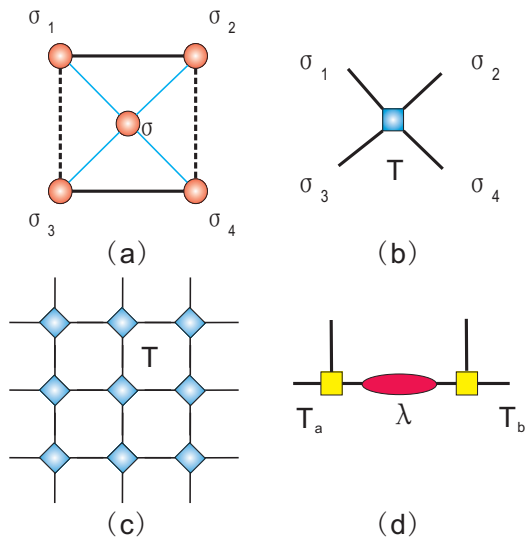}
\caption{(Color online) Tensor network representation in LTRG calculations of this present system. (a) the cell lattice for tensor construction; (b) the schematic representation of tensor $\mathbf{T}$; (c) the tensor network of the partition function; and (d) the matrix product state (Ref. [\onlinecite{Ltrg}]) obtained by contracting the tensor network in (c). }
\label{Tensor}
\end{figure}

In addition, by making use of the elements in the $\lambda$ matrix in the matrix product state (MPS), shown in Fig. \ref{Tensor} (d), obtained by the contraction of the tensor network in Fig. \ref{Tensor} (c), we can introduce the quasi-entanglement entropy following the similar way in one-dimensional quantum lattice system
\begin{equation}\label{entropy}
  S_Q=-\Sigma_i \lambda_i log\lambda_i.
\end{equation}
Actually, the MPS obtained by contracting the tensor network is equivalent to the ground state of its corresponding quantum 1D model \cite{GNT}. The singular point of the quasi-entanglement entropy which corresponds to the phase transition in quantum 1D model can be used to determine $T_c$ or $T_{c,TK}$ for the 2D classical lattice. The temperature dependence of the quasi-entanglement entropy $S_Q$ at $\beta=1.6$ is calculated, as presented in Fig. \ref{c and e} (b), where a sharp peak of  $S_Q$ is seen at the temperature that is the same as the temperature at which the susceptibility exhibits a singularity. It appears that the quasi-entanglement entropy defined by Eq. (\ref{entropy}) can be used to detect the KT phase transition.

\begin{figure}[tbp]
\includegraphics[width = 1.0\linewidth,clip]{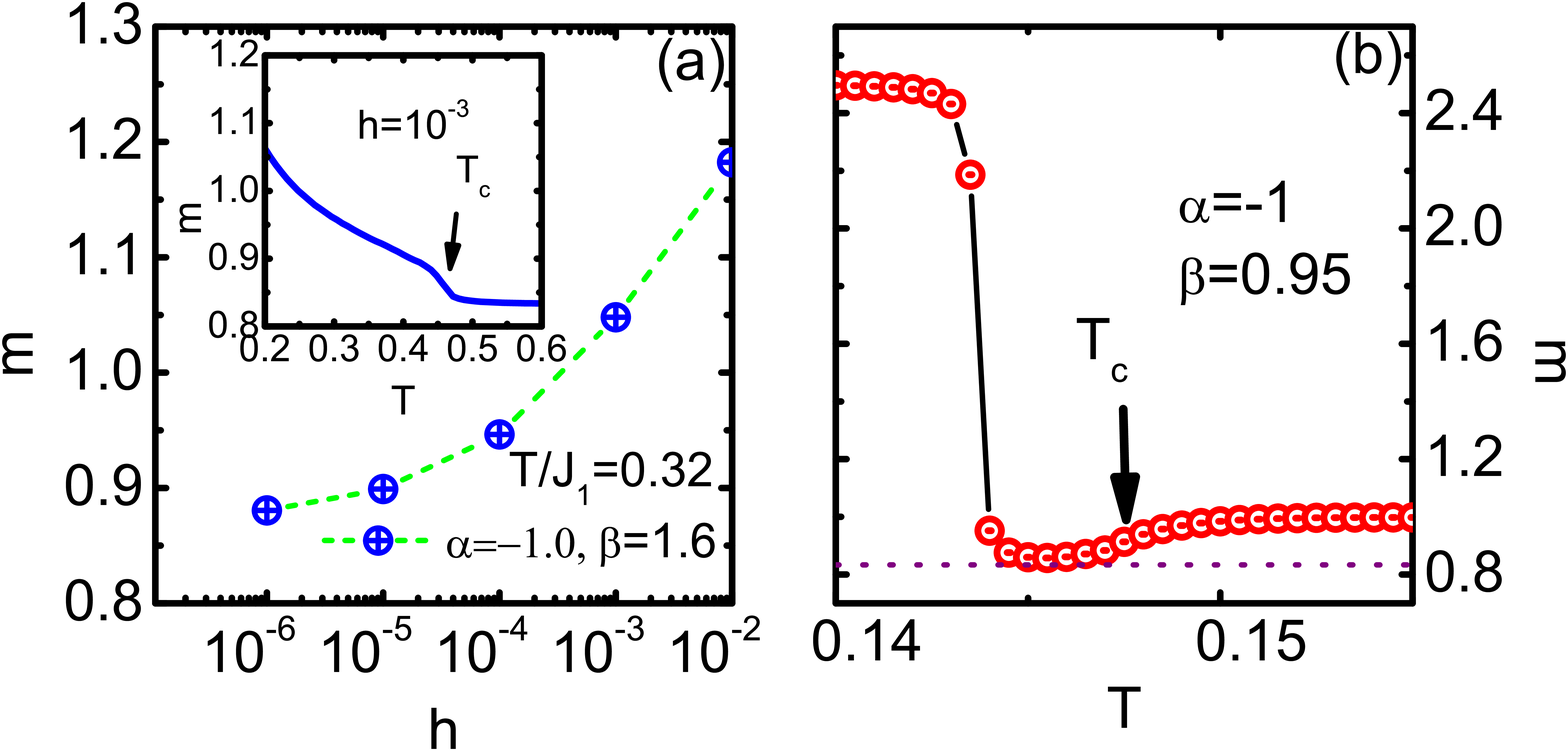}
\caption{(Color online) (a) The magnetization versus small magnetic field at temperature $T/|J_1|=0.32$. The inset is the magnetization versus temperature in a small field $h=10^{-3}$. Here $\beta = 1.6$. (b) Temperature dependence of $m$ in the reentrance region.}
\label{m_scaling}
\end{figure}

\section{Reentrant Phenomena and Phase Diagram}

The KT phase transition implies the quenching of the partial order below $T_{c,KT}$ that further supports the disordered ground state when $\beta$ $>$ 1 owing to the highly degeneracies in the ground state and the strong fluctuations in the Potts model. Under $T_{c,KT}$, the floating phase is very sensitive to external perturbations. Fig. \ref{m_scaling} (a) gives the induced $m$ by a tiny field ($\sim 10^{-3}$) for $\beta=1.6$, where one may see that $m$ increases fast with increasing the small field, showing the character of the floating phase. The inset of Fig. \ref{m_scaling} (a) demonstrates that $m$ decreases with temperature in a small field, and at a certain temperature, has a rapid drop to the value of $5/6$, indicating again that it is in the floating state.

From the above discussions, we come to conclude that the anisotropic 3-state Potts model on the generalized Kagom\'e lattice for $\beta>1$ has the topological phase transition of the KT type. An interesting question then arises: can a reentrant phase exist under a phase without a partial order through the KT transition? Let us answer this question below.

\begin{figure}[tbp]
\includegraphics[width = 0.8\linewidth,clip]{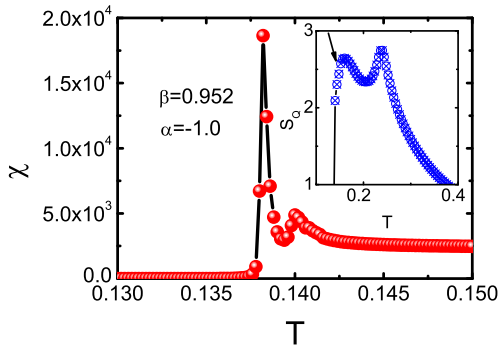}
\caption{(Color online) Temperature dependence of the susceptibility of the anisotropic 3-state Potts model on the generalized Kagom\'e lattice in the reentrance region. The inset shows the corresponding quasi entanglement entropy.}
\label{Sus}
\end{figure}

To search for the reentrant region, we would collect the information from the specific heat,  susceptibility and the quasi-entanglement entropy to determine the phase diagrams. We first begin with $\alpha = -1$ and take $\beta$ as the variable. In calculations we determine the critical temperature of ferromagnetic phase transition by means of the peak of the specific heat and the KT phase transition temperature according to the peak of the quasi-entanglement entropy. By performing a large-scale calculations, we find that the parameter range of $\beta$  where the reentrant phenomenon occurs is in between $[0.94, 1]$, and in this small region the system is most frustrated. Fig. \ref{m_scaling}(b) gives the temperature dependence of magnetization for $\alpha = -1$ and $\beta=0.95$, and it can be seen that with increasing temperature the magnetization decreases sharply to the value of $5/6$, after experiencing the flat change, then remarkably increases, suggesting that there should be two phase transitions in this case, namely, one from F state to paramagnetic (P) state, and another from P state to floating state.  Such transitions can also be clearly seen from the susceptibility, as shown in Fig. \ref{Sus}, where the temperature dependence of the susceptibility is presented for $\beta$ = 0.952 and $\alpha = -1$. There are two peaks: the first peak at $T_{c,f}$ $\approx$ 0.136 corresponds to the phase transition from F to P phase, and the second peak at the critical temperature $T_{c,xi}$ $\approx$ 0.140 indicates the phase transition from P to $X$ phase. The inset of Fig. \ref{Sus} gives the temperature dependence of the quasi-entanglement entropy, where three peaks are observed: the two peaks at the critical temperatures 0.136 and 0.140 are too close to be visibly separated, which just correspond to the two peaks of susceptibility, while the third peak occurs at $T_{c,xf}$$\approx$ 0.242, indicating the phase transition from X phase to P phase. Therefore, for $\beta$ = 0.952, with increasing temperature, the system undergoes phase transitions from P to X to P phase, which is nothing but the reentrant phenomenon.

\begin{figure}[tbp]
\includegraphics[width = 0.8\linewidth]{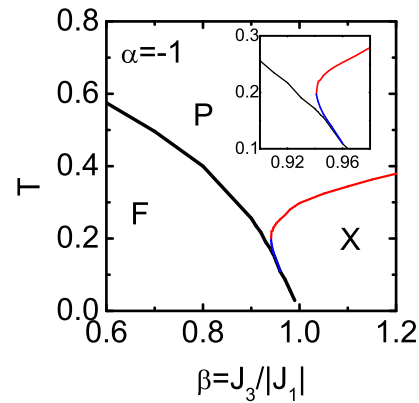}
\caption{(Color online) The phase diagram of the anisotropic 3-state Potts model on the generalized Kagom\'e lattice with $\alpha$ = -1. The inset indicates the reentrant region. F: ferromagnetic phase; P: paramagnetic phase; X: Floating phase.}
\label{Phase1}
\end{figure}

The whole phase diagram for $\alpha=-1$ is depicted in Fig. \ref{Phase1}, where three phases including F, P and X phases are identified. One may note that there exists a small region $\beta \in [0.94, 1]$ where the reentrance occurs, as shown in the inset of Fig. \ref{Phase1}. It can be observed that in this region and with increasing temperature, the system goes first into the P phase from the F phase, and then enters into the X phase, and then goes into again the P phase. In addition, some further computations reveal that the region of $\beta$ for which the reentrant phenomenon can appear is proportional to $|\alpha|$ ($\alpha < 0$) and shrinks to zero at $\alpha$ = 0 that is just the typical Kagom\'e lattice. It is also worth mentioning that the phase diagram in Fig. \ref{Phase1} is similar to that of the generalized Kagom\'e Ising model \cite{reent}. Since in the 3-state Potts model the thermal fluctuations are stronger than in the Ising model, the phase boundaries in Fig. \ref{Phase1} stand lower than in the diagram of Ref. [\onlinecite{reent}]. In the latter diagram the region of the partial ordered state $A$ is replaced by the floating phase labeled by $X$.

\begin{figure}[tbp]
\includegraphics[width = 0.8\linewidth,clip]{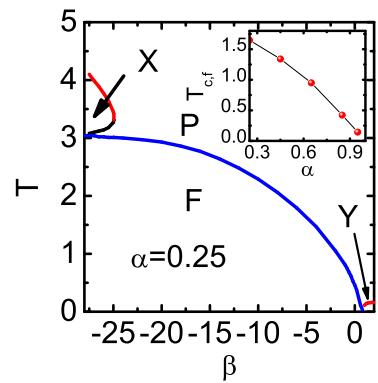}
\caption{(Color online) Phase diagram of the anisotropic 3-state Potts model on the generalized Kagom\'e lattice with 0 $<$ $\alpha$ $<$ 1. F: Ferromagnetic; P: Paramagnetic; X and Y: floating phases. The inset is $\alpha$-dependence of the ferromagnetic phase transition critical temperature $T_{c,f}$.}
\label{Phase2}
\end{figure}

When 0 $<$ $\alpha$ $<$ 1, there is also a reentrant property, as shown in Fig. \ref{Phase2}. In the phase diagram with $\alpha$ = 0.25, there are four phases, including F, P, X and Y phases, where Y is also a floating phase appearing in a very small region when $\beta > 0.8$. There exists a reentrant region between F and X phases when $\beta$ $<$ -24.94. This region, determined by the same procedure as above, is the same as the generalized Kagom\'e Ising model \cite{reent,book}, but the critical temperature $T_{c,f}$ is lowered than that in the Ising case due to strong thermal fluctuations. It is found that the area of X phase increases with the value of $\alpha$ in the cost of the decreased scope of the F phase. It is clear that in the inset of Fig. \ref{Phase2} the ferromagnetic phase transition temperature $T_{c,f}$ goes to zero when $\alpha$ increases near to 1.  The finite temperature phase transition from X to P phase is also of the KT type, and the critical temperature is obtained through the calculation of the entanglement entropy or susceptibility.

\begin{figure}[tbp]
\includegraphics[width = 0.8\linewidth,clip]{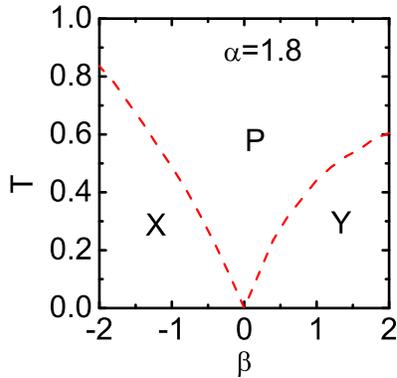}
\caption{(Color online) Phase diagram of the anisotropic 3-state Potts model on the generalized Kagom\'e lattice with $\alpha$ $>$ 1. P: Paramagnetic; X and Y: floating phases. }
\label{phase3}
\end{figure}

When $\alpha$ $>1$ the F phase disappears, and there will be only two floating phases $X$ and $Y$ in the phase diagram, as shown in Fig. \ref{phase3}. In  this case, no reentrant phenomenon occurs. In addition, the partial ordered state with configuration $B$ in the Ising model (Ref. \onlinecite{reent}) when $\alpha$, $\beta$ $>$ 0 is vanished due to highly thermal fluctuations in the 3-state Potts model as in Fig. \ref{Phase2}. Besides, we found through a number of calculations that when $q$ $>$ 3, the $q$-state Potts model on the generalized Kagom\'e lattice exhibits no any reentrant property.

\section{summary}

 To summarize, we have studied the anisotropic 3-state Potts model on the generalized Kagom\'e lattice by utilizing the LTRG method. The phase diagrams in the plane of temperature versus $\beta$ for the cases of $\alpha <0$,  $0<\alpha <1$ and  $\alpha >1$ are obtained. Different phases including ferromagnetic, paramagnetic and floating phases are identified. The phase boundaries are determined by observing the singularities in the specific heat, magnetization, susceptibility and quasi-entanglement entropy. For the two cases of $\alpha <0$ and $0<\alpha <1$, the reentrant phenomena are observed in small regions, respectively. By studying the behaviors of the correlation length, we found that the phase transition for $\beta>1$ is of the KT type, in sharp contrast to the case of $\beta<1$ where the correlation length has a quite different behavior from that of $\beta>1$. 
 
 By comparing with the generalized Kagom\'e Ising model, where the partial ordered  phases exist, the present 3-state Potts model possesses floating phases $X$ and $Y$ owing to strong thermal fluctuations.
 Such floating phases do not have any local order parameters.  Through the present study, we may remark that the appearance of the reentrant phenomena in 2D classical lattices is mainly caused by  frustrations as well as the geometrical structure of the lattices, which has also close relation with the freedom $q$ of the local spin. Importantly, the existence of the partial ordered state in the ground state may not be the necessary condition for the reentrant phenomena.

\acknowledgments

We are grateful to Shou-Shu Gong, Zheng-Chuan Wang and Qing-Rong Zheng for useful discussions. This work is supported in part by the NSFC (Grant No. 90922033 and No. 10934008), the MOST of China (Grant No. 2012CB932901, No. 2013CB933401), and the CAS.

\end{document}